\documentstyle[amssymb,aps,graphicx,epsf,twocolumn]{revtex}

\begin{document}
\draft
\title{Magnetic Field Induced Gap and Kink Behavior of Thermal Conductivity}
\author{E.~J. Ferrer$^{1}$, V.~P. Gusynin$^{2}$$^*$, and V. de la Incera$^{1} $}
\address{$^{1}$ Physics Department, SUNY-Fredonia, Fredonia, NY 14063, USA}
\address{$^2$Department of Physics, Nagoya University, Nagoya 464-8602, Japan\\
{\rm (\today)}}
\address{~\\
\parbox{14cm}{\rm \medskip
The thermal conductivity of a quasiparticle (QP) system  described by a
relativistic four-fermion interaction model in
the presence of an external magnetic field is calculated. It is
shown that, for narrow width of quasiparticles, the thermal
conductivity, as a function of the applied magnetic field,
exhibits a kink behavior at a critical field $B_{c}\sim T^{2}$.
The kink is due to the opening of a gap in the QP spectrum at a critical magnetic field
$B_c$ and to the enhancement of the transitions between the zeroth
and first Landau levels. Possible applications of the results are discussed.
}}
\maketitle

\narrowtext

It is now well known from the study of field theoretical models
that an external magnetic field can be a strong catalyst for a
symmetry breaking leading to the generation of a fermion dynamical
mass even at the weakest attractive interactions \cite{prl94}.
This phenomenon of magnetic catalysis (MC) of symmetry breaking
has been shown to be essentially model independent and to have
wide applications in several physical systems \cite
{prl94,qed4,Klimenko,misc}, in particular, in condensed matter
physics \cite {Nick,Nick2,Semenoff,Liu}. The essence of the MC
effect lies in the dimensional reduction of the fermion pairing
dynamics when the pairing energy is much less than the Landau gap
$\sqrt{eB}$ ($B$ is the magnitude of the magnetic field
induction). In this case, the fermions are mostly confined to the
lowest Landau level (LL) and their attractive interaction,
regardless how small it may be, leads to mass generation.

In recent years, several authors \cite{Nick2,Semenoff,Liu}
suggested that the
magnetic catalysis may play a relevant role in the physics of high-$%
T_{c} $ superconductors in external magnetic fields. This idea was
inspired in part by the outcome of various experiments
\cite{Krishana,Aubin,Ando} initiated by Krishana's et al.
 on the thermal conductivity of high-$%
T_{c} $ compounds in the presence of an external magnetic field
perpendicularly applied to the cuprate plane. According to these
experiments, at temperatures significantly lower than the critical
temperature of superconductivity, the profile of the thermal
conductivity $\kappa $ versus the magnetic field displays a sharp
break in its slope at a transition field $B_{c}$, followed then by
a plateau region in which it ceases to change with increasing
field up to the highest attainable fields $\sim 14T$.

To understand how the magnetic catalysis was related
\cite{Nick2,Semenoff,Liu} to these
experimental observations, let us recall that high-$%
T_{c} $ superconducting cuprates are d-wave superconductors which
implies that their Fermi surface is characterized by nodal points
where the gap function vanishes. Then, the low-energy
quasiparticle (QP) dynamics is concentrated mainly around these
nodal points because these states can be occupied even at very low
temperatures. The QP excitations in the vicinity of each node can
be described by a massless Dirac Lagrangian
\cite{condmatdirac,Durst-Lee}. Hence, a natural candidate to
modelling the QP interactions in a d-wave superconductor is a
$3$-dimensional relativistic four-fermion interaction
(Nambu-Jona-Lasinio (NJL) type) model. It is well known \cite
{prl94,Klimenko} that the gap equations of such a kind of models
in the presence of an external magnetic field lead to the
generation of a dynamical fermion mass
(QP gap), which scales (at zero temperature) with the field as $m\sim \sqrt{%
eB}$. The critical temperature at which the dynamical mass
vanishes is determined by the dynamical mass at zero temperature,
and therefore, it scales with the magnetic field with the same law
as the zero-temperature mass. One of the outcomes of the above
mentioned experiments \cite{Krishana} was that the critical
temperature for the appearance of the kink-like behavior of the
thermal conductivity scales with the magnetic field as $T_{c}\sim
\sqrt{B}$. Then, the square root field dependence of the critical
temperature in $(2+1)-$dimensional NJL models was considered as a
hint that the MC phenomenon could be important to explain the
experimental results of Krishana et al. \cite{Krishana}.

By now, several possible mechanisms for the arising plateau have
been proposed \cite{Nick2,Semenoff,Liu,Balatsky,Laughlin,Franz}.
All of them are based on the adoption of the QP picture as the
low-energy excitations at the nodes of the $d-$wave symmetric
order parameter. In the approach of Refs.\cite
{Nick2,Semenoff,Liu}, following the ideas of Ref.
\cite{Gorkov,Anderson}, it was assumed that the QP spectrum in the
magnetic field was characterized by Landau levels. Furthermore, in
Ref. \cite{Nick2,Semenoff,Liu,Laughlin} it was supposed that a QP
gap opens at the critical field $B_{c}$ leading to the exponential
vanishing of the quasiparticle contribution in the thermal
conductivity. Since the total conductivity is assumed to be the
sum of the QP term $\kappa _{{\rm QP}}$ and the phonon term
$\kappa _{{\rm ph}}$ with the phononic part independent of the
field, this would lead to a plateau for high fields. Laughlin
\cite {Laughlin} relates the QP gap to a weakly first order phase
transition leading to the development of an additional small
parity-violating superconducting order parameter. In another line
of reasoning \cite {Nick2,Semenoff,Liu,Balatsky}, the drop in the
QP's conductivity, hence the plateau in the total conductivity, is
ascribed to the opening of a field-induced energy gap in $d$-wave
superconductors due to a second order phase transition.

On the contrary, the approach of Franz \cite{Franz} does not appeal to the
generation of any QP gap. Albeit the phononic part of the thermal
conductivity is also considered to be field-independent, in Franz' scenario
the QPs conductivity itself becomes field-independent above a crossover
field $B_{c}$. The effect is primarily due to the compensation of the
enhancement of the QP density of states, arising from the Doppler shift of
QP's spectrum in the superfluid velocity field around vortices (Volovik
effect \cite{doppler}), and the growth of QP's width caused by the
scattering on vortices. The Volovik effect certainly plays a decisive role
at weak magnetic fields and low temperatures (less than $1$ K), where the
increase in the thermal conductivity was predicted theoretically \cite
{Kubert}, and later confirmed in the experiment \cite{Aubin}, to follow the $%
\sqrt{B}$ dependence of the density of states. However, Volovik's
arguments are based on semiclassical calculations valid in case
one has well isolated vortices with supercurrents circulating
around them. At higher fields, when vortices begin to overlap,
this semiclassical picture should be replaced by a more adequate
quantum mechanical treatment.

It should be stressed that none of the above works address the
question of the kink itself in the thermal conductivity. Moreover,
we should mention that the applicability of the Landau level
quantization scheme used in the QP picture of the d-wave
superconductor in a constant magnetic field has been recently
subjected to intense criticism by several authors
\cite{Melnikov,Tesa,F-T-approach}. The Landau level-like spectrum
proposed by Gorkov and Schiffer \cite{Gorkov}, and Anderson
\cite{Anderson} was based on the assumption that the spatially
dependent superfluid velocity can be neglected, but according to
Melnikov \cite{Melnikov}, the superfluidity velocity strongly
mixes individual Landau levels. In order to take into account the
effects of spatially varying supercurrents, Franz and Tesanovich
\cite{Tesa} introduced a singular gauge transformation to trade
the phase of the order parameter for a ficticious $U(1)$ gauge
field. As a result, the two fluxes, due to the external magnetic
field and the $U(1)$ field, cancel each other on average, washing
out the whole picture of Landau level quantization. In this
framework the low-energy quasiparticles are in fact Bloch waves
described by massless Dirac fermions moving in a vector potential
of physical supercurrents but zero average magnetic field. While
the arising scenario might seem promising, the approach of Ref.
\cite{Tesa,F-T-approach} itself is not free of problems: there is
no good foundation from a physical point of view to use singular
gauge transformations and to introduce fictitious gauge fields,
both of which are needed in this scheme. Moreover, because of the
singular transformations, the physical picture itself becomes
strongly dependent on the way such transformations are implemented
(\textsl{i.e.}, different singular gauge transformations can give
rise to different spectra, depending on the regularization scheme
\cite{F-T-approach}). Finally, as far as we know, it has not been
proved up to now that such a formulation can be applicable in the
regime of weak-to-moderate magnetic fields, where the individual
vortices overlap and the magnetic field can be considered to be
fairly uniform.

The aim of the present paper is to discuss a mechanism that
generates the kink effect within the framework of the MC
phenomenon. We emphasize that although, as shown in the
derivations that follow, our numerical curves are in close
qualitative agreement with the experimental curves of Ref.
\cite{Krishana}, we do not pretend to claim that our results
explain the experimental findings of Krishana et. al.
\cite{Krishana}, as our calculations are based on the Landau level
quantization whose applicability to d-wave superconductivity, as
mentioned above, is not well established. Our main result in this
paper is then of a more general and theoretical character. That
is, we show that the MC phenomenon can be responsible for a kink
effect in the thermal conductivity of (2+1)-dimensional
relativistic fermion systems, and we point out the very important
finding that the kink-like effect close to the critical field is
essentially model independent, since it is determined by the
critical behavior of the dynamically generated mass near the phase
transition point. This fact makes our result relevant beyond the
particular model under consideration, linking it to the
universality class of theories with such a critical behavior.

 In connection with this, we would like to call the
attention of the reader to a new class of recently discovered
materials \cite{graphite}, which could be a playground for
applications of the phenomenon of magnetic catalysis. The
so-called  highly oriented pyrolitic graphites (HOPG) have layered
structure with two isolated points in the Brillouin zone where the
dispersion is linear, so that the electronic states can be
described here too in terms of relativistic Dirac particles
\cite{Gonzales}. It was already suggested \cite{K-2} that the
magnetic catalysis in the system of Coulomb-interacting Dirac
fermions can provide an explanation for the semimetal-insulator
phase transition observed in HOPG in the presence of a magnetic
field perpendicular to the layers. Since the quasiparticles in a
graphite do not suffer from the Bloch waves versus Landau levels
dilemma, our present results should have full strength there and
we anticipate that the thermal conductivity of these systems
should exhibit a behavior similar to the one we are reporting
here.

In this paper we present for the first time a consistent
calculation of the thermal conductivity in the presence of an
external magnetic field in a model with the simplest four-fermion
interaction for quasiparticles. We use the same constant magnetic
field approximation that was already explored in
Ref.\cite{Nick2,Semenoff} to calculate the thermal conductivity.
However, our calculation deviates considerably from what was done
in \cite{Nick2,Semenoff}. Not only we take into account the
contribution of all Landau levels, but the definition of the heat
current itself is different. In the limit of narrow width of
quasiparticles ($\Gamma \ll T,\sqrt{B}$), after a gap is opened,
the thermal conductivity exhibits a new term proportional to
$\sigma ^{2}$ ($\sigma $ is the gap). This term originates from
the compensation of the matrix elements of transitions between the
zeroth and the first LLs (behaving as $1/B$) and the LL density of
states which in turn is proportional to $B$. This is one more
manifestation of the MC phenomenon: not only a gap is induced by
the magnetic field, but the transitions between the zeroth and the
first LLs are enhanced. In mean-field
approximation and near the phase transition point the gap behaves like $%
\sigma \simeq 0.523\sqrt{eB-eB_{c}}$ what yields a positive contribution
into the slope of the thermal conductivity leading to a jump (kink effect)
of $\kappa (B)$ at $B=B_{c}$.

We start from the following NJL model in an external magnetic
field in $(2+1)$ dimensions,
\begin{equation}
\hspace{-1.5mm}{\cal L}=\bar{\psi}_{a}[i\gamma ^{0}\hbar \frac{\partial }{%
\partial t}+v_{D}\gamma ^{i}(i\hbar \frac{\partial }{\partial x^{i}}-\frac{e%
}{c}A_{i})]\psi _{a}+\frac{g}{2N}(\bar{\psi}_{a}\psi _{a})^{2},
\label{lagrangian}
\end{equation}
In (\ref{lagrangian}) the vector potential is taken in the symmetric gauge $%
A_{\mu }=\left( 0,-\frac{B}{2}x_{2},\frac{B}{2}x_{1}\right) $, and
$v_{D}$ is a characteristic velocity \cite{V}. In what follows we
take $\hbar =v_{D}=k_{B}=1$ and absorb $c$ into the charge $e$. We
will restore these constants when needed. We assume also that the
fermions carry an additional flavor index $a=1,\dots ,N$ ($N=2$
for realistic $d$-wave superconductors). The Dirac $\gamma $
matrices are taken in a reducible four-component representation.

The Lagrangian density (\ref{lagrangian}) is invariant under the discrete
(chiral) symmetry $\psi \rightarrow \gamma _{5}\psi ,\,\bar{\psi}\rightarrow
-\bar{\psi}\gamma _{5},$ which forbids the fermion mass generation in
perturbation theory. The mass generation can be studied introducing an
auxiliary field $\sigma =-(g/N)\bar{\psi}_{i}\psi _{i}$ by means of the
Hubbard-Stratanovich trick which permits one to integrate over fermion
fields in the path integral representation of the partition function. The
field $\sigma $ has no dynamics at the tree level but it acquires the
kinetic term due to fermion loops. Likewise, a nontrivial minimum of the
effective potential (the expectation value of $\sigma $) gives mass to
fermions and spontaneously breaks the discrete symmetry leading to a neutral
condensate of fermion-antifermion pairs. Studying the minimum of the
effective potential we find that, at a fixed temperature $T$, there is a
critical value of the magnetic field $\sqrt{eB_{c}}/T\simeq 4.148$ such that
for subcritical fields $eB\leq eB_{c}$ the gap is zero, while for $eB>eB_{c}$
a nontrivial gap appears. The critical curve equation has the form $%
(v_{D}/c)^{2}10^{10}B=21.5\cdot T^{2}$ for $B$ measured in Tesla.
Using the approximated value of the characteristic velocity
$v_{D}\simeq\ 10^{7}cm/s $  \cite{Liu,K-2}, we obtain the critical curve $%
B=0.014\cdot T^{2}$. Notice that the obtained critical curve fits well the experimental curve of Ref.\cite{Krishana}%
.

To derive an expression for the static thermal conductivity in an isotropic
system, we follow the familiar linear response method and apply Kubo's
formula \cite{Ambegaokar}
\begin{equation}
\kappa =-\frac{1}{TV}{\rm Im}\int\limits_{0}^{\infty }dtt\int
d^{2}x_{1}d^{2}x_{2}\langle u_{i}(x_{1},0)u_{i}(x_{2},t)\rangle ,
\label{heat_cur_cor}
\end{equation}
where $V$ is the volume of the system and $u_{i}(x,t)$ is the heat-current
density operator. The brackets denote averaging in the canonical ensemble.
Physically, the thermal conductivity $\kappa $ appears as a coefficient in
the equation relating the heat current to the temperature gradient ${\vec{u}}%
=-\kappa {\vec{\nabla}}T$ under the condition of absence of particle flow.
If we neglect the chemical potential, the heat density coincides with the
energy density $\epsilon $, hence the quantity ${\vec{u}}$ that satisfies
the continuity equation $\dot{\epsilon}(x)+{\vec{\nabla}.}{\vec{u}(x)}=0$
can be interpreted as the heat current. From the Lagrangian density (\ref
{lagrangian}) we find $u_{i}=\frac{i}{2}\left( \bar{\psi}\gamma _{i}\partial
_{0}\psi -\partial _{0}\bar{\psi}\gamma _{i}\psi \right) $.

Neglecting vertex corrections \cite{Lee} the calculation of the thermal
conductivity reduces to the evaluation of the bubble diagram \cite{Langer}.
In this approximation, making use of the spectral representation for the
fermion Green's function, we arrive at the following expression for the
thermal conductivity in the Matsubara formalism
\begin{equation}
\kappa =\frac{1}{32\pi T^{2}}\hspace{-1.5mm}\int\limits_{-\infty }^{\infty }%
\hspace{-1.5mm}\frac{d\omega \omega ^{2}}{\cosh ^{2}\frac{\omega }{2T}}%
\hspace{-1mm}\int \hspace{-1mm}d^{2}k\,{\rm tr}\hspace{-1mm}\left[ \gamma
^{i}A(\omega ,{\vec{k}})\gamma ^{i}A(\omega ,{\vec{k}})\right] .
\label{kappathroughA}
\end{equation}
The spectral function $A(\omega ,{\vec{k}})=-\pi ^{-1}{\rm Im}S^{R}(\omega
+i\varepsilon ,{\vec{k}})$ is derived from the fermion propagator in a
magnetic field decomposed over Landau level poles \cite{prl94,Chodos},
\begin{eqnarray}
&&A(\omega ,{\vec{k}})=e^{-\frac{{\vec{k}}^{2}}{eB}}\frac{\Gamma }{2\pi }%
\hspace{-0.5mm}\sum\limits_{n=0}^{\infty }\hspace{-0.3mm}\frac{(-1)^{n}}{%
M_{n}}\left[ \frac{(\gamma ^{0}M_{n}+\sigma )f_{1}({\vec{k}})+\hspace{-0.5mm}%
f_{2}({\vec{k}})}{(\omega -M_{n})^{2}+\Gamma ^{2}}\right.  \nonumber \\
&&+\left. \frac{(\gamma ^{0}M_{n}-\sigma )f_{1}({\vec{k}})-f_{2}({\vec{k}})}{%
(\omega +M_{n})^{2}+\Gamma ^{2}}\right] ,\,M_{n}=\sqrt{\sigma ^{2}+2eBn}, \\
&&f_{1}({\vec{k}})=2\left[ P_{-}L_{n}(s)-P_{+}L_{n-1}(s)\right] ,f_{2}({\vec{%
k}})=4({\vec{k}\cdot }{\vec{\gamma})}L_{n-1}^{1}(s).  \nonumber
\end{eqnarray}
Here $P_{\pm }=(1\pm i\gamma ^{1}\gamma ^{2})/2$ are projector operators, $%
L_{n},L_{n}^{1}$ are Laguerre's polynomials ($L_{-1}^{1}=0$), $s=2{\vec{k}}%
^{2}/{eB,}$ and $\sigma $ is the dynamical fermion mass obtained from the
finite temperature gap equation in the magnetic field. We introduced also
the quasiparticles width replacing $\varepsilon $ in the definition of $%
A(\omega ,{\vec{k}})$ by finite $\Gamma $, which is due to interaction
processes, in particular, scatterings on impurities. Straightforward
calculation of the trace and further integration over momenta in Eq. (\ref
{kappathroughA}) produce Kronecker's delta symbols $\delta _{n,m-1}+\delta
_{m,n-1}$ due to orthogonality of Laguerre's polynomials. This implies that
only those transitions between neighbor Landau levels contribute to the heat
transfer. After performing one of the sums in (\ref{kappathroughA}) we
obtain
\begin{eqnarray}
\kappa &=&\frac{eB\Gamma ^{2}N}{\pi ^{2}T^{2}}\sum\limits_{n=0}^{\infty
}\int\limits_{0}^{\infty }\frac{d\omega \omega ^{2}}{\cosh ^{2}\frac{\omega
}{2T}}\frac{1}{(\omega ^{2}+M_{n}^{2}+\Gamma ^{2})^{2}-4\omega ^{2}M_{n}^{2}}
\nonumber \\
&\times &\frac{(\omega ^{2}+M_{n}^{2}+\Gamma ^{2})(\omega
^{2}+M_{n+1}^{2}+\Gamma ^{2})-4\omega ^{2}\sigma ^{2}}{[(\omega
^{2}+M_{n+1}^{2}+\Gamma ^{2})^{2}-4\omega ^{2}M_{n+1}^{2}]}.  \label{kappa}
\end{eqnarray}
Further summation over $n$ in Eq. (\ref{kappa}) can be carried out expanding
the integrand in terms of partial fractions. The resulting sums are
expressed through digamma functions and the final expression for $\kappa $
is written as follows
\begin{eqnarray}
&&\kappa =\frac{N\Gamma ^{2}}{2\pi ^{2}T^{2}}\hspace{-0.5mm}%
\int\limits_{0}^{\infty }\hspace{-1mm}\frac{d\omega \omega ^{2}}{\cosh ^{2}%
\frac{\omega }{2T}}\frac{1}{(eB)^{2}+(2\omega \Gamma )^{2}}\left\{ 2\omega
^{2}\right. +  \nonumber \\
&&\left. \frac{(\omega ^{2}+\sigma ^{2}+\Gamma ^{2})(eB)^{2}-2\omega
^{2}(\omega ^{2}-\sigma ^{2}+\Gamma ^{2})eB}{(\omega ^{2}-\sigma ^{2}-\Gamma
^{2})^{2}+4\omega ^{2}\Gamma ^{2}}\,-\right.  \nonumber \\
&&\left. \frac{\omega (\omega ^{2}-\sigma ^{2}+\Gamma ^{2})}{\Gamma }{\rm Im}%
\psi \left( \frac{\sigma ^{2}+\Gamma ^{2}-\omega ^{2}-2i\omega \Gamma }{2eB}%
\right) \right\} .  \label{kappamaggeneral}
\end{eqnarray}
This formula is the main result of our paper and we can use it now to study
the different asymptotic regimes. It is easy to show that in the limit of
zero field $\kappa $ reduces to
\begin{equation}
\kappa _{0}=\frac{N}{4\pi ^{2}T^{2}}\int\limits_{0}^{\infty }\frac{d\omega
\omega ^{2}}{\cosh ^{2}\frac{\omega }{2T}}\left[ 1+\frac{\omega ^{2}+\Gamma
^{2}}{\omega \Gamma }\arctan \frac{\omega }{\Gamma }\right] ,
\label{thermcondfree}
\end{equation}
where we put $\sigma =0,$ since the mass is not generated in the zero-field
weak coupling NJL model. Eq. (\ref{thermcondfree}), up to an overall factor $%
k_{B}^{2}(v_{F}^{2}+v_{\Delta }^{2})/\hbar v_{F}v_{\Delta }$, coincides with
the corresponding expression obtained in Ref.\cite{Franz}. With the overall
factor replacing $N$ ($=2$ in real $d-$wave superconductor), Eq. (\ref
{thermcondfree}) reproduces the universal (or residual) thermal conductivity
at low $T$ in the so-called ``dirty'' limit, $T\ll \Gamma $\thinspace \cite
{Lee}. The residual conductivity was recently observed in experiments \cite
{univ-kappa_exp} giving explicit confirmation of the existence of gapless
quasiparticles in $d$-wave cuprates at $T<T_{c}$.

Let us consider the case $\Gamma \ll T$ with $B\neq 0$. In this
approximation we obtain
\begin{equation}
\kappa \simeq \frac{N\Gamma }{4\pi T^{2}}\left\{ \frac{\sigma ^{2}}{\cosh
^{2}\frac{\sigma }{2T}}+4\sum\limits_{n=1}^{\infty }\frac{n(\sigma
^{2}+2eBn) }{\cosh ^{2}\frac{\sqrt{\sigma ^{2}+2eBn}}{2T}}\right\} .
\label{kappa_large_B}
\end{equation}
Asymptotically, at $\sqrt{eB}\gg \sqrt{eB_{c}}\simeq T$, the dynamical mass
behaves as $\sigma \sim \sqrt{eB}$ what leads to an exponential decrease as
in the case of gapless fermions. However, the term $\cosh ^{-2}\sigma /2T$
in Eq. (\ref{kappa_large_B}) is of order one when $\sigma \alt 2T$, thus,
there is no suppression of this term for a certain range of fields where it
is almost field independent (plateau region).

Near the phase transition point $\sigma \simeq a\sqrt{eB-eB_{c}}$
and the first term in Eq. (\ref{kappa_large_B}) gives positive
contribution to the slope of the thermal conductivity at $eB\ge
eB_{c}$ leading to the jump in the slope of $\kappa $ (kink
effect) at $eB=eB_{c}$. The parameter $a$ is model-dependent, but
the scaling of $\sigma$ is determined by the universality class of
the theory. For the NJL model (\ref{lagrangian}) we find, in
mean-field approximation, $a\simeq 0.523$.

\begin{figure}
\epsfxsize=9cm \epsfysize=7cm\epsffile{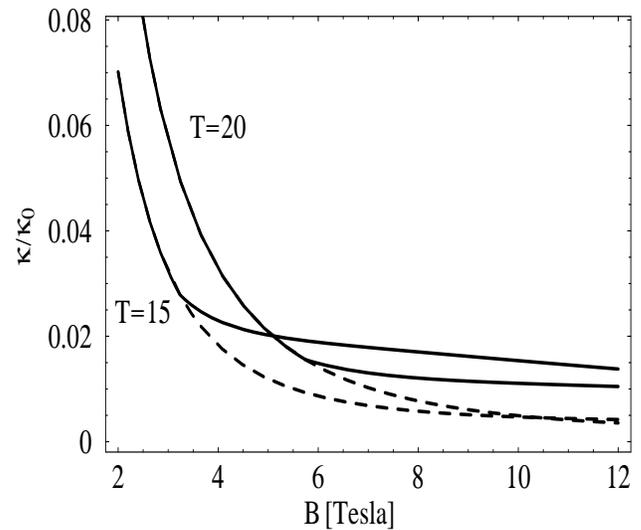}
 \caption{The magnetic
field dependence of $\kappa $ at $T=20 K$ and $T=15 K$ in the
narrow width case $(\Gamma=5 K).$ The solid lines represent
$\kappa
/\kappa _{0}$ when a QP gap $\sigma $ is MC-induced at $B\geq B_{c}(T)$ ($%
B_{c}(20)=5.75{\rm T},\,B_{c}(15)=3.23{\rm T}$). The dashed lines
represent the behavior of $\kappa /\kappa _{0}$ when $\sigma $
remains zero at $B\geq B_{c}(T)$. } \label{fig:1}
\end{figure}

The explicit appearance of the kink has been corroborated by numerical
calculations as shown in Fig.\ref{fig:1}. Notice the break in the slope of $%
\kappa $ (kink effect) at the critical value $B_{c}$ in the presence of $%
\sigma $. For $B>B_{c}$ the kink is followed by a region where $\kappa $ is
only weakly dependent on the field. While decreasing the temperature, the
position of the kink moves to the left in accordance with the critical line $%
B_{c}=0.014T^{2}$. Notice the similarity of our results (Fig.
\ref{fig:1}) with the experimental behavior reported in
\cite{Krishana} for high-Tc superconductors. Moreover, Fig.
\ref{fig:1} also reproduces the experimentally observed crossing
of the curves that occurs with decreasing $T$ in such a way that
the lower $T$ curve reaches the higher value at large fields.

Because of the important role played by the first term in (\ref
{kappa_large_B}) to get the kink effect, it is instructive to clarify its
origin. From Eq. (\ref{kappa}) one can see that while the contribution of
transitions between Landau levels with $n\ge 1$ in the integrand behaves as $%
1/(eB)^{2}$ for large fields, the contribution of transitions between the
zeroth and the first LLs decreases only as the first power of the field ($%
\sim 1/eB$). Since the density of LLs is proportional to $eB$, this implies
that the transitions between the zeroth and the first LLs are not
suppressed, even though the gap between the levels grows with the field.
Note that for gapless QPs the transitions with $n=0$ are completely
suppressed in the regime $\Gamma \to 0$: indeed, their contribution is
proportional to $\delta (\omega )$ and the integrand in (\ref{kappa})
contains an $\omega ^{2}$ factor. For gapped QPs, $\delta (\omega )$ is
replaced by $\delta (\omega \pm \sigma )$ and the transitions with $n=0$
survive.

In summary, in the present paper the thermal conductivity at
$B\neq 0$ in the framework of a relativistic four-fermion model
was calculated. Assuming a uniform magnetic field approximation,
we showed that the thermal conductivity exhibits a kink behavior
when the field reaches the critical value $B_{c}=0.014\cdot
T^{2}$. Two main features determine this effect: the generation of
a QP gap in a magnetic field (MC phenomenon) and the lack of
suppression of zeroth-first LLs transitions. We stress that the
appearance of a kink in the thermal conductivity is model
independent, being only determined by the critical behavior of the
induced dynamical mass near the phase transition. The universality
character of this result leaves an open window for possible future
applications.

We would like to acknowledge V.A.~Miransky for stimulating discussions and
careful reading of the manuscript, and A.~Hams and I.~Shovkovy for help in
numerical calculation. This research has been supported in part by NSF under
Grants PHY-0070986 (E.J.F. and V.I.), PHY-9722059 (E.J.F., V.P.G. and V.I.)
and POWRE-PHY-9973708 (V.I.). The work of V.P.G. is supported also by the
SCOPES-project 7UKPJ062150.00/1 of the Swiss NSF and Grant-in-Aid of Japan
Society for the Promotion of Science (JSPS) \#11695030. V.P.G. wishes to
acknowledge JSPS for financial support.

\begin{references}
\bibitem[*]{}  On leave of absence from Bogolyubov Institute for Theoretical
Physics, Kiev, 03143 Ukraine.

\bibitem{prl94}  V.P.~Gusynin, V.A.~Miransky, and I.A.~Shovkovy, \prl
{\bf 73}, 3499 (1994); 
\prd {\bf 52}, 4718 (1995). 

\bibitem{qed4}  V.P.~Gusynin, V.A.~Miransky, and I.A.~Shovkovy, \pl B {\bf %
349}, 477 (1995); 
\prd {\bf 52}, 4747 (1995); 
\prl {\bf 83}, 1291 (1999). 

\bibitem{Klimenko}  K.G.~Klimenko, Z. Phys. C{\bf 54}, 323 (1992).

\bibitem{misc}  C.N. Leung, Y.J. Ng, and A.W. Ackley, \prd {\bf 54}, 4181
(1996); 
D.K. Hong, Y. Kim, and S.-J. Sin, \prd {\bf 54}, 7879 (1996);
D.S. Lee, C.N. Leung, and Y.J. Ng, \prd {\bf 55}, 6504 (1997);
V.P.~Gusynin and I.A.~Shovkovy, Phys. Rev. D {\bf 56}, 5251 (1997).
E.J. Ferrer and V. de la Incera, Int. J. Mod. Phys., A {\bf 14}, 3963
(1999); 
Phys. Lett. B {\bf 481}, 287 (2000); 
E.J. Ferrer, V.P. Gusynin, and V. de la Incera, Phys. Lett. B {\bf 455}, 217
(1999); 
G.W. Semenoff, I.A. Shovkovy, and L.C.R. Wijewardhana, Phys. Rev. D {\bf 60}
, 105024 (1999); 
V.Ch. Zhukovsky {\sl et al.}, hep-th/0012256.

\bibitem{Nick}  K. Farakos and N.E. Mavromatos, cond-mat/9710288. 

\bibitem{Nick2}  K.~Farakos, G.~Koutsoumbas, and N.E.~Mavromatos, Int. J.
Mod. Phys. B{\bf 12}, 2475 (1998). 

\bibitem{Semenoff}  G.W.~Semenoff, I.A.~Shovkovy, and L.C.R.~Wijewardhana,
Mod. Phys. Lett. A{\bf 13}, 1143 (1998). 

\bibitem{Liu}  W.V.~Liu, Nucl. Phys. B {\bf 556}, 563 (1999).

\bibitem{Krishana} K. Krishana \textsl{et al.}, Science, \textbf{277}, 83
(1997);
 N. P. Ong {et al.}, cond-mat/9904160.

\bibitem{Aubin}  H.~Aubin {\sl et al.}, Phys. Rev. Lett., {\bf 82}, 624
(1999). 

\bibitem{Ando}  The kink and plateau observed in zero-field cooled BSCCO
\cite{Krishana,Aubin} have not been found in the field-cooled
BSCCO samples [Y. Ando {\sl et al.}, cond-mat/9812265] meaning
that more refined measurements are needed to clarify the status of
Krishana's {\sl et al.}
data.

\bibitem{condmatdirac}  G.W.~Semenoff and L.C.R.~Wijewardhana, \prl
{\bf 63}, 2633 (1989); 
J.B.~Marston, Phys. Rev. Lett. {\bf 64}, 1166 (1990).
N.~Dorey and N.E.~Mavromatos, Nucl. Phys. B {\bf 368}, 614 (1992);
S.H.~Simon and P.A.~Lee, \prl {\bf 78}, 1548 (1997).

\bibitem{Durst-Lee} A. C. Durst and P. A. Lee,Phys. Rev. B {\bf 62}, 1270
(2000).

\bibitem{Balatsky}  A.V. Balatsky, Phys. Rev. Lett. {\bf 80}, 1972 (1998).

\bibitem{Laughlin}  R.B.~Laughlin, \prl  {\bf 80}, 5188 (1998),
T.V.~Ramakrishnan, J. Phys. Chem. Solids, {\bf 59}, 1750 (1998).

\bibitem{Franz}  M.~Franz, Phys. Rev. Lett. {\bf 82}, 1760 (1999).

\bibitem{Gorkov}  L.P.~Gor'kov and J.R.~Schrieffer, Phys. Rev. Lett. {\bf 80},
3360 (1998); 

\bibitem{Anderson}  P.W.~Anderson, cond-mat/9812063;

\bibitem{doppler}  G.E.~Volovik, JETP Lett.,{\bf 58}, 469 (1993).

\bibitem{Kubert}  C.~Kubert and P.J.~Hirschfeld, Phys. Rev. Lett., {\bf 80},
4963 (1998). 

\bibitem{Melnikov}A.~ S.~ Mel'nikov J. Phys. Cond. Matter {\bf 11},
4219 (1999).

\bibitem{Tesa}M.~Franz and Z.~Tesanovic, Phys. Rev. Lett. {\bf 84},
554 (2000).

\bibitem{F-T-approach}L.~Marinelli et al., Phys. Rev. B {\bf 62}, 3499 (2000);
J.~Ye, Phys. Rev. Lett., {\bf 86}, 316 (2001);
O.~Vafek et al, Phys. Rev. B {\bf 63}, 134509 (2001);
J.~Ye and A.~Millis, cond-mat/0101032.
D.~Knapp, C.~Kallin and A. J.~Berlinsky, Phys. Rev. B, 64, 014502
(2001);
A. ~Vishwanath, Phys. Rev. Lett. 87, 217004 (2001).

\bibitem{graphite} M.~S.~ Sercheli et al., cond-mat/0106232.

\bibitem{Gonzales}G. Semenoff, Phys. Rev. Lett. \textbf{53}, 2449
(1984); F.~D.~M.~ Haldane,  \textsl{ibid} \textbf{61}, 2015
(1988); J.~Gonzales, F.~Guinea, and M.~A.~H.~Vozmediano, Phys.
Rev. Lett. {\bf77}, 3589 (1996).

\bibitem{K-2}  D. V. Khveshchenko, Phys. Rev. Lett. \textbf{87}, 206401
(2001).

\bibitem{V} In the QP picture of the d-wave superconductor $v_D$ was taken
as $v_{D}=\sqrt{v_{F}v_{\Delta }}$ with $v_{F},v_{\Delta }$ being
the velocities perpendicular and tangential to the Fermi surface
respectively. They originate from the quasiparticle excitation
spectrum in the
vicinity of the gap nodes that takes the form of an anisotropic Dirac cone $E(k)=\sqrt{%
v_{F}^{2}k_{1}^{2}+v_{\Delta }^{2}k_{2}^{2}}$. After rescaling
coordinates this leads to Eq. (\ref{lagrangian}). In the Dirac
picture of the quasiparticle excitations in layered graphite,
$v_D$ is proportional to the width of the electronic $\pi$-orbital
band.



\bibitem{Ambegaokar}  V.~Ambegaokar and A.~Griffin, Phys. Rev. {\bf 137},
1151 (1965). 

\bibitem{Lee}  It was argued that for small impurity densities the thermal
conductivity is unaffected by vertex corrections, see
\cite{Durst-Lee}.

\bibitem{Langer}  J.S. Langer, Phys. Rev. {\bf 127}, 5 (1962);
V. Ambegaokar and L. Tewordt, Phys. Rev. {\bf 134}, A805 (1964).

\bibitem{Chodos}  A.~Chodos, K.~Everding, and D.A.~Owen, Phys. Rev. D {\bf 42%
}, 2881 (1990). 

\bibitem{univ-kappa_exp}  L.~Taillefer et al., Phys. Rev. Lett. {\bf 79},
483 (1997); 
K.~Behnia et al., J. Low Temp. Phys. {\bf 117}, 1089 (1999).
\end{references}
\end{document}